\documentclass[12pt]{article}

\usepackage{amsmath}
\usepackage{amssymb}
\usepackage{amsthm}
\usepackage{upref}

\title{\bf A Continuous Model of Computation}
\author{J.F. Traub}
\date{Department of Computer Science \\
Columbia University \\
New York, NY 10027 \\ 
September, 1998}

\begin{document}

\maketitle

\begin{abstract}
Although the Turing-machine model of computation is widely
used in computer science it is fundamentally inadequate as a
foundation for the theory of modern scientific computation.
The real-number model is described as an alternative. Physicists
often choose continuous mathematical models for problems ranging
from the dynamical systems of classical physics to the operator
equations and path integrals of quantum mechanics.These mathematical
models use the real or complex number fields and we argue that the
real-number model of computation should be used in the study of the
computational complexity of continuous mathematical models.

The study of continuous complexity is called information-based
complexity. In this expository article we apply information-based
complexity to topics such as breaking the curse of dimensionality,
approximating the calculation of path integrals, and solving ill-posed
problems. Precise formulations of these ideas may be found in
J. F. Traub and A. G. Werschulz, \lq\lq Complexity and Information\rq\rq,
Cambridge University Press, 1998. 
\end{abstract}

\vskip 2pc


A central dogma of computer science is that the Turing machine model 
is \underline{the} appropriate abstraction of a digital computer. 
Physicists who've thought about this also seem to favor the Turing 
machine model. For example, Roger Penrose  \cite{1} devotes some 60 pages to a description of this abstract model of computation and its implications. I will argue that physicists should consider the real-number model of computation as more appropriate and useful for scientific computation.

First, I'll introduce the four \lq\lq worlds\rq\rq\ that will play a role; see Figure 1. Above the horizontal line are two real worlds; the world of physical phenomena and the computer world, where simulations are performed. Below the horizontal line are two formal worlds; a mathematical model of the physical phenomenon and a model of computation which is an abstraction of a physical 
computer. {\it We get to choose both the mathematical model and the model of computation. What type of models should we choose?}

\vskip 2pc
\begin{center}
\begin{tabular}{|c|c|} \hline
Real-World Phenomena & Computer Simulation   \\ \hline
Mathematical Model   & Model of Computation  \\ \hline
\multicolumn{2}{c}{ } \\
\multicolumn{2}{c}{\bf Figure 1. Four worlds}
\end{tabular}
\end{center}
\vskip 1pc

The mathematical model, which is often continuous, is chosen by the physicist. Continuous models range from the dynamical systems of classical physics to the operator equations and path integrals of quantum mechanics. That is, it would take an infinite number of bits to represent a single real number; an infinite number of bits are not available in the universe. Real numbers are utilized because they are a powerful and useful construct. Let us accept that today continuous models are widely used in mathematical physics and that they will continue to occupy that role for at least the foreseeable future. But the computer is a finite state machine. 
{\it What should we do when the continuous mathematical model meets the 
finite state machine?}

In the next Section I will compare and contrast two models of computation: the Turing machine and real-number models. In the interest of full disclosure I want to tell you that I've always used the real number model in my work but will do my best to present balanced arguments.

I will then show how the real-number model is used in the study of the computational complexity of continuous mathematical models. (Computational complexity is a measure of the minimal computational resources required to solve a mathematically posed problem.) This is the province of a branch of complexity theory called information-based complexity or IBC for brevity. After providing  a formulation of IBC I'll try to demonstrate the power of this theory.

\vskip 2pc
\section*{\bf Pros and Cons of Two Models of Computation}
\vskip 1pc

Although many readers are familiar with the Turing machine model I will 
describe it very briefly. Then, after describing the real-number model, 
I will discuss the pros and cons of these two models.

Alan Turing, one of the intellectual giants of the twentieth century, 
defined his machine model to establish the unsolvability of Hilbert's 
Entscheidungsproblem \cite{2}.
The Turing machine model has a tape which is of infinite length. The tape is divided into a squares each of which is either blank or contains a single mark. For any particular case the input, calculation, and output are finite. That is, beyond a certain point the tape is blank. There is a reading head which reads one square at a time and after each operation moves one square to the left or right.  The machine has a finite number of internal states. Given the initial state and input sequence on the tape the machine changes its state and the reading head prints a symbol and moves one square. Finally, the machine decides when to halt.

We turn to a very brief description of the real-number model; a precise formulation may be found in the literature \cite{3,4}.
The crux of this model is that we can store and perform arithmetic operations and comparisons on real numbers exactly and at unit cost. We can also perform information operations but since these are harder to understand without a motivating example we defer describing them until the next Section. They are not essential for our discussion here.

The real-number model has a long history. Alexander Ostrowski used it in his work on the computational complexity of polynomial evaluation in 1954. I used the real-number model for research on optimal iteration theory in 1964. Shmuel Winograd and V\"{o}lker Strassen used the real-number model in their work on algebraic complexity in the late sixties. Henryk Wo\'{z}niakowski and I used it in our 1980 monograph on information-based complexity. In 1989, Lenore Blum, Michael Schub, and Steven Smale provided a formalization of the real-number model for continuous combinatorial complexity and established the existence of NP-complete problems over the reals. Their work led to a surge of research on computation over the reals.

Both these models of computation are abstractions of real digital computers. Writing about the Turing machine model Penrose \cite{1}
 states \lq\lq It is the unlimited nature of the input, 
calculation space, and output which tells us that we are considering only a mathematical idealization.\rq\rq\ 
Which abstraction to use depends on how useful that abstraction is for certain purposes.
What are the pros and cons of these two models of computation? I'll begin with the pros of the Turing machine model. It is desirable to use a finite-state abstraction of a finite-state machine. Moreover, the Turing machine's simplicity and economy of description are attractive. Another plus is that it is universal. It is universal in two senses. The first is the Church-Turing thesis, which states that what a Turing machine can compute may be considered a universal definition of computability. (Computability on a Turing machine is equivalent to computability in Church's lambda calculus.) Of course, one cannot prove this thesis; it appeals to our intuitive notion of computability. It is universal in a second sense. All \lq\lq reasonable\rq\rq\ 
machines are polynomially equivalent to Turing machines. 
Informally, this means that if the minimal time to compute an output on a Turing machine is $T(n)$ for an input of size $n$  
and if the minimal time to compute an output on any other machine is $S(n)$,
then $T(n)$ and $S(n)$ are polynomially related. 
Therefore, one might as well use the Turing machine as the model of 
computation.

I'm not convinced by the assertion that all reasonable machines are polynomially equivalent to Turing machines, but I'll defer my critique for the cons of the Turing machine. See Table 1 for a summary of the pros of the Turing machine model.

\vskip 2pc
\begin{center}
\begin{itemize}
\item Desirable to use a finite-state model for a finite-state machine
\item Universal 
\begin{itemize}
\item Church-Turing thesis
\item All \lq\lq reasonable\rq\rq\ machines are polynomially equivalent to Turing 
machines
\end{itemize}
\end{itemize}
\vskip 1pc
Table 1. Pros of Turing machine model 
\end{center}
\vskip 1pc

I'll turn to the cons of the Turing machine model. I believe it is not natural to use this discrete model in conjunction with continuous mathematical models. Furthermore, estimated running times on a Turing machine are not predictive of scientific computation on digital computers. One reason for this is that scientific computation is usually done with fixed precision floating point arithmetic. The cost of arithmetic operations is independent of the size of the operands. Turing machine operations depend on number size.
Finally, there are interesting computational models which are not polynomially equivalent to a Turing machine. Consider the example of a UMRAM. The acronym reveals the important properties of this model of computation. It is a Random Access Machine where Multiplication is a basic operation and memory access and multiplication and addition can be performed at a Unit cost. This seems like a reasonable abstraction of a digital computer since multiplication and addition on fixed-precision floating point numbers cost about the same. But the UMRAM is not polynomially equivalent to a Turing machine! However, a RAM, which does not have multiplication as a fundamental operation is polynomially equivalent to a Turing machine. The cons of the Turing machine are summarized in Table 2.

\vskip 2pc
\begin{center}
\begin{itemize}
\item Not natural to use a discrete model of computation in conjunction with the continuous models of physics
\item  Not predictive of running time of a scientific computation on a digital computer
\item Not all \lq\lq reasonable\rq\rq\ machines are equivalent to Turing machines
\end{itemize}
\vskip 1pc
Table 2. Cons of the Turing machine model
\end{center}
\vskip 1pc

I now turn to the pros of the real-number model. As I've stated before, the mathematical models of physics are often continuous and use real (and complex) numbers. That is, physicists assume a continuum. It seems natural to use the real numbers in analyzing the numerical solution of continuous models on a digital computer.

Most scientific computation uses fixed-precision floating point arithmetic. Leaving aside the possibility of numerical instability, computational complexity in the real number model is the same as for fixed-precision floating point. Therefore the real-number model is predictive of running times for scientific computations.  Studying the computational complexity in the real-number model has led to new superior methods for doing a variety of scientific calculations.

A third reason for using the real-number model is that it permits the full power of continuous mathematics. We will see an example in the final Section when I discuss a result on non-computable numbers and its possible implications for physical theories. Using Turing machines the results takes a substantial part of a monograph to prove. With analysis, the analogous result in the real number model is established in about a page.

The argument for using the power of analysis was already made in 1948 by John von Neumann, one of the leading mathematical physicists of the century and a father of the digital computer. In his Hixon Symposium lecture \cite{5}, 
von Neumann argues for a \lq\lq more specifically analytical theory of automata and information.\rq\rq\ 
He writes: \lq\lq There exists today a very elaborate system of formal logic, and specifically, of logic as applied to mathematics. This is a discipline with many good sides, but also serious weaknesses ... Everybody who has worked in formal logic will confirm that this is one of the technically most refractory parts of mathematics. The reason for this is that it deals with rigid, all-or-none concepts, and has very little contact with the continuous concept of the real or of the complex number, that is, with mathematical analysis. Yet analysis is the technically most successful and best-elaborated part of mathematics ... The theory of automata, of the digital, all-or-none type as discussed up to now, is certainly a chapter in formal logic.\rq\rq\ These observations may be used mutatis mutandis as an argument for the real number model.

Blum, Felipe Cucker, Shub, Smale \cite{6} argue for the real-number model stating \lq\lq The point of view of this book is that the Turing model ... 
is fundamentally inadequate for giving such a foundation to the theory of modern scientific computation, where most of the algorithms ...  
\underline{are real number algorithms}.\rq\rq\ 
The pros of the real-number model are summarized in Table~3.

\vskip 2pc
\begin{center}
\begin{itemize}
\item \lq\lq Natural\rq\rq\ for continuous mathematical models
\item Predictive of computer performance on scientific problems
\item Utilizes the power of continuous mathematics
\end{itemize}
\vskip 1pc
Table 3. The pros of the real-number model
\end{center}
\vskip 1pc

A con of the real-number model is that digital representations of real numbers does not exist in the real world. Even a single real number would require infinite resources to represent exactly. Thus the real-number model is not finitistic. The Turing machine model is also not finitistic since it utilizes an unbounded tape. It is therefore {\it potentially} infinite. Thus, to paraphrase George Orwell, the Turing machine model is less infinite than the real-number model. It would be attractive to have a finite model of computation (the Turing machine model is discrete but unbounded). There are finite models, such as circuit models and linear bounded automata, but they are special-purpose. The cons of the real-number model are given in Table 4.

\vskip 2pc
\begin{center}
\begin{itemize}
\item It is impossible to construct a physical device that implements the real-number model
\item It is preferable to use a finite-state abstraction of a finite-state machine
\end{itemize}
\vskip 1pc
Table 4. Cons of real-number model
\end{center}

\vskip 2pc
\section*{\bf The Formulation of Information-Based Complexity}
\vskip 1pc

To see the real-number model in action I'll indicate how to formalize computational complexity issues for continuous mathematical problems and then, in the following Section, I'll describe a small selection of recent results. I'll use the example of $d$-dimensional integration to motivate concepts and because of its importance in fields ranging from physics to mathematical finance. In the next Section I'll touch briefly on the case $d=\infty$, that is, path integrals.

We wish to compute the integral of a real-valued function, $f$, of $d$
 variables over the unit cube in $d$ dimensions. Typically, we have to settle for computing a numerical approximation with an error $\epsilon$. 
To guarantee an $\epsilon$-approximation we have to know some global information about the integrand. We assume that the class of integrands has smoothness $r$. One way of defining such a class is to let $F_r$ consist of those functions whose derivatives of order through $r$ satisfy a uniform bound.

A real function of a real variable cannot be entered into a digital computer. We evaluate $f$ at a finite number of points and we call the set of values of $f$ the local information, for brevity information about $f$. An algorithm combines the function values into a number which approximates the integral.

In the worst case setting we guarantee an error at most $\epsilon$ for every
$f\in F$. The computational complexity, for brevity complexity, is the least cost of computing the integral to within $\epsilon$ for every $f\in F$. 
We charge unity for every arithmetic operation and $c$ comparison and  for every function evaluation. Typically $c \gg 1$ . I want to stress that the complexity depends on the problem and $\epsilon$ but not on the algorithm. Every possible algorithm, whether or not it is known, and all possible points at which the integrand is evaluated are permitted to compete when we consider least possible cost.

It can be shown that if $F=F_r$, then the complexity of our integration problem is of order $\epsilon^{-d/r}$. If $r=0$, e.g., if our set of integrands consists of uniformly bounded continuous functions, the complexity is infinite. That is, it is impossible to solve the problem to within $\epsilon$. Let $r$ be positive and in particular let $r=1$. Then the complexity is of order $\epsilon^{-d}$. Because of the exponential dependence on $d$ we say the problem is computationally intractable. This is sometimes called the curse of dimensionality. Very large $d$ occur in practice and in the next Section we'll see a problem from mathematical finance where $d=360$.

I'll compare this $d$-dimensional integration problem with the well-known Traveling Salesman Problem (TSP), an example of a discrete combinatorial problem. The input is the location of the  cities and the desired output is the minimal route; the city locations are usually represented by a finite number of lists. Therefore the input can be exactly entered into a digital computer. The complexity of this problem is unknown but conjectured to be exponential in the number of cities. That is, the problem is conjectured to be computationally intractable and many other combinatorial problems are conjectured to be intractable. We summarize some differences between the integration and TSP problems in Table 5.

\vskip 2pc
\begin{center}
\begin{tabular}{p{2.5in}p{2.5in}}
\multicolumn{1}{c}{TSP}  & \multicolumn{1}{c}{Integration} \\ [.2in]
$\bullet$ Information about the input is exact &
   $\bullet$ Information about the input is partial \\ [.1in]
$\bullet$ Conjectured intractable in the number of cities &
   $\bullet$ Proven intractable in number of variables \\ [.1in]
$\bullet$ Model of computation is usually Turing machine or random 
access machine &
   $\bullet$ Model of computation is usually real-number
\end{tabular}
\vskip 1pc
Table 5. Comparison of TSP and integration
\end{center}
\vskip 1pc

Most problems in scientific computation which involve multivariate functions belonging to $F_r$ have been proven computationally intractable in the number of variables in the worst case setting. These include partial differential equations \cite{7}, nonlinear optimization \cite{8}, nonlinear equations \cite{9}, 
approximation \cite{10}, and integral equations \cite{7}. 
The assumption $f\in F_r$ is important. For example, if convexity is assumed, nonlinear optimization requires only on the order of $\log(\epsilon^{-1})$ 
function evaluations \cite{8}.

In general, IBC is defined by the assumptions that the information concerning the mathematical model is

\begin{center}
\begin{tabular}{l}
$\bullet$  partial \\
$\bullet$  contaminated \\
$\bullet$  priced
\end{tabular}
\end{center}

\noindent
Referring to the integration example the mathematical input is the integrand and the information is a finite set of function values. It is partial because the integral cannot be recovered from function values. For a partial differential equation the mathematical input is the functions specifying the initial value and/or boundary conditions. Generally, the mathematical input is replaced using a finite number of information operations. These operations may be functionals on the mathematical input or physical measurements which are fed into a mathematical model. We can now state as an additional specification of the real-number model that information operations are permitted at cost $c$ \cite{3,4}.

In addition to being partial the information is often contaminated \cite{11} 
by, for example, round-off or measurement error. If the information is partial or contaminated it is impossible to solve the problem exactly. Finally, the information is priced. As examples, functions can be costly to evaluate or information needed for oil exploration models is obtained by setting off shocks. With the exception of certain finite-dimensional problems, such as roots of systems of polynomial equations and problems in numerical linear algebra, the problems typically encountered in scientific computation have information that is partial and/or contaminated and priced.

IBC theory is developed over linear abstract spaces such as Banach and Hilbert spaces; the applications are typically to multivariate functions. As part of our study of complexity we investigate optimal algorithms, that is, an algorithm whose cost is equal or close to the complexity of the problem. This has sometimes led to new solution methods.

The reason that we can often obtain the complexity and an optimal algorithm for IBC problems is that partial and/or contaminated information permits arguments at the information level. This level does not exist for combinatorial problems where we usually have to settle for trying to establish a complexity hierarchy and conjectures such as ${\rm P}\ne {\rm NP}$.

A powerful tool at the information level is the notion of the radius of information, $R$. The radius of information measures the intrinsic uncertainty of solving a problem using given information. 
We can compute an $\epsilon$-approximation if and only if $R\leq \epsilon$. 
The radius depends only on the problem being solved and the available information; it is independent of the algorithm. The radius of information is defined in all IBC settings; we've seen the worst case setting above and will meet two additional settings in the next Section. The radius can be used to define the 
{\it value of information} \cite{4} for continuous problems which agrees with the mutual information as obtained from the continuous analogue of Shannon entropy for some important cases but disagrees for other. The value of information seems superior in the cases where the two measures differ.

\vskip 2pc
\section*{\bf Recent Advances in Information-Based Complexity}
\vskip 1pc

I'll present a small selection of recent IBC advances: complexity of high dimensional integration, complexity of path integrals, and whether ill-posed problems are solvable.

Continuous multivariate problems are typically intractable in dimension for the worst case setting. There are two ways to attempt to break the curse of dimensionality. We can weaken the worst case guarantee to a stochastic assurance or we can change the class of mathematical inputs. For high-dimensional integrals we will see both strategies at play.

Recall that in the worst case setting the complexity of $d$-dimensional integration is of order $\epsilon^{-d/r}$ . 
As is widely known, the expected cost of Monte Carlo (MC) is of order $\epsilon^{-2}$, independent of $d$ . 
(This is equivalent to the expected error of MC is of order $n^{-2}$ 
 if $n$ evaluations are used.) This holds even if the class of integrands is, say, continuous and uniformly bounded; that is, if $r=0$. 
But there is no free lunch. MC carries only a stochastic assurance of small error. 

A second stochastic setting is the average case setting. In contrast with the randomized setting where MC permitted, this is a deterministic setting with an a priori measure on the space of mathematical inputs. The guarantee is only that the expected error with respect to this measure is at most $\epsilon$ 
and the complexity is the least expected cost.

What is the complexity of integration in the average case setting and where should the integrand be evaluated? This problem was open for some 20 years until it was solved by Wo\'{z}niakowski  in 1991. 
He assumed that $r=0$  and that the measure on the space of integrands is the Wiener sheet measure. He proved that the average complexity is of order
$\epsilon^{-1}\left(\log \epsilon^{-1} \right)^{(d-1)/2}$
and that the integrand should be evaluated at the points of a low discrepancy sequence (LDS). 
Discrepancy theory is a much studied branch of number theory and many LDS are known. Roughly speaking, a set of points has low discrepancy if a rather small number of points is distributed as uniformly as possible in the unit cube in 
$d$ dimensions.

The integral is approximated by 
$\sum_{i=1}^n a_i f(t_i)$. 
If the $t_1$ are chosen from a random distribution, this is MC; if the 
$t_i$ are a LDS it is called quasi-Monte Carlo (QMC). Ignoring the polylog 
factor, $\left( \log \epsilon^{-1}\right)^{(d-1)/2}$,
in Wo\'{z}niakowski's theorem the cost of QMC is $\epsilon^{-1}$ which is much smaller than the cost $\epsilon^{-2}$ of MC, especially if the error $\epsilon$ 
is small. Hence there has been much interest in QMC. The polylog factor in 
Wo\'{z}niakowski's theorem and a similar factor which occurs in the Koksma-Hlawka inequality is innocent if $d$ is modest. But in certain applications $d$ is large. For example, in a Collateralized Mortgage Obligation (CMO) calculation $d=360$.
For such a value of $d$  the polylog factor looks ominous.

Computer experiments \cite{12} by Spassimir Paskov with the CMO and other financial instruments led to the surprising conclusion that QMC always beat MC, often by orders of magnitude. Anargyros Papageorgiou and I have shown that using generalized Faure points, which is a particular LDS sequence, QMC achieves accuracy of $10^{-2}$, 
which is sufficient for finance, with just $170$ function evaluations.
A recent paper by Ian Sloan and Wo\'{z}niakowski \cite{13} may provide an explanation. Their paper is based on the observation that many problems of mathematical finance are highly non-isotropic; that is, some dimensions are much more important than  others. This is due, at least in part, to the discounted value of future money. They prove the existence of a quasi-Monte Carlo method 
whose \underline{worst case cost} is at most of order $\epsilon^{-p}$  with $p\leq 2$.

There is a certain resemblance between this sequence of events and physics. There were beautiful and widely accepted results in discrepancy theory. They failed to correctly predict the out come of certain numerical experiments. This led to the development of a new theory. Although the Sloan-Wo\'{z}niakowski theory looks very promising, it leaves some open questions and research into why QMC performs so well for high-dimensional integrals of mathematical finance is currently a very active area.

QMC looks promising for problems besides those occurring in mathematical finance. For example, Papageorgiou and I \cite{14} reported test results on a model integration problem of Brad Keister \cite{15} suggested by integration problems in physics. Again QMC outperformed MC by a large factor, for $d$ as large as a hundred. The cost of MC remains of order $\epsilon^{-2}$ ; 
an analysis by Papageorgiou has shown that the cost of QMC is of order 
$\epsilon^{-1} \left(\log \epsilon^{-1}\right)^{1/2}$.

We end this discussion with a caveat. There are many problems for which it has been established that randomization does not break intractability. One instance is function approximation \cite{10}. The following question has been open for many years. Characterize those problems for which randomization breaks intractability.
A new area of research is the complexity of path integrals which are defined by
$$S(f)=\int_X f(x) \, \mu(dx)$$
where $\mu$ is a probability measure on an infinite-dimensional space $X$. 
Numerous applications occur in quantum mechanics, continuous financial mathematics, and the solution of partial differential equations. We wish to compute an $\epsilon$-approximation for all $f\in F$. The usual method of calculating path integrals is to approximate the infinite-dimensional integral by one of finite dimension which is calculated by Monte Carlo. The first paper \cite{16} on the complexity of calculating path integrals is by Greg Wasilkowski and 
Wo\'{z}niakowski who assume the measure is Gaussian.

Here I'll briefly describe very recent work of Leszek Plaskota, 
Wasilkowski and Wo\'{z}niakowski on a new algorithm for certain Feyman-Kac path integrals which concern the solution of the heat equation
$$\frac{\partial z(u,t)}{\partial t} = \frac 12 \frac{\partial^2 z(u,t)}
{\partial u^2} + v(u)z(u,t), \; z(u,0)=v(u)$$
with the Feyman-Kac integral of the form
$$z(u,t)=\int_C v(x(t)+u) e^{\int_0^t V(x(s)+u)\, ds}\, w(dx).$$
Here $t>0$, $V$ is a potential function, $v$ is an initial condition function,
$C$ is the space of continuous functions and $w$ is the Wiener measure. Assume that $v=1$  and that $V$ is four times continuously differentiable. Then Alexandre Chorin's algorithm, which uses MC, gives a stochastic assurance of error 
at most $\epsilon$ at cost of order $\epsilon^{-2.5}$.

The new algorithm is deterministic and the assurance is worst case. Its cost 
is of order $\epsilon^{-.25}$. Thus the new algorithm enjoys the advantage of a worst case guarantee at much lower cost. It, however, suffers from a drawback in that it requires the calculation of certain coefficients given by weighted integrals. Under some circumstances one may view the computation of these as a precomputation cost which is not counted in the cost of the algorithm.

The cost of the new algorithm gives an upper bound on the complexity. A lower bound has also been established and for certain classes $G$  the two bounds are essentially the same. Then the complexity of this Feyman-Kac path integral is known and the new algorithm is essentially optimal. The results are more general than I've indicated here; I want to give the reader a taste of this work.

As my last example I'll discuss ill-posed problems from the viewpoint of computability theory and IBC to show the power of analysis at work. Marian Pour-El and Ian Richards \cite{17} established what Penrose \cite{1} characterizes as 
\lq\lq a rather startling result.\rq\rq\ 
Consider a wave equation for which the data provided at an initial time determines the solution at all other times. Then there exist computable initial conditions with the property that at a later time the value of this field is non-computable. A similar result holds for the backward heat equation. These are very particular instances of the following theorem. An unbounded linear transformation can take computable inputs into non-computable outputs. Both the examples mentioned above are examples of unbounded linear transformations.

Pour-El and Richards devote a large part of their monograph to prove this result using computability theory. An analogous result using IBC over the reals was established by Arthur Werschulz \cite{18}. Werschulz's proof utilizes the power of analysis and is about one page long. His approach has several other advantages as we shall see.

I remind the reader of the classical theory of ill-posed problems due to Jacques Hadamard. Roughly speaking, a problem is ill-posed if small changes in the problem specification can case large changes in the problem solution. As one can gather from his terminology Hadamard felt there was something inherently wrong  with trying to solve ill-posed problems. Given an ill-posed problem, one should attempts to reformulate it, so it can be correctly set. However, there are many important ill-posed problems such as remote sensing and computational vision, that occur in practice. 

Assume we want to compute $Lf$  where $L$ is a linear operator. The operator is said to be unbounded if $\sup \| Lf\|=\infty $ for $\| f\| \leq 1$. It is well known that $Lf$ is ill-posed if and only if $L$ is unbounded. Examples of problems whose solution is given by an unbounded linear operator are the wave equation with an initial condition which is not twice differentiable and the backward heat equation. Although Werschulz's result holds on normed linear spaces, for simplicity I'll describe it for function spaces. He assumes that the function $f$
 (in the case of differential equations this might be the initial condition) cannot be entered  into a  digital computer. He discretizes $f$ by evaluating it at a discrete number of points. Werschulz proves that if the problem is ill-posed it is impossible to compute an $\epsilon$-approximation to the solution at finite cost even for arbitrary large $\epsilon$. Thus the problem is unsolvable. Note that this is a much stronger result than non-computability.

But the best is yet to come. Since the definition of unbounded linear operator involves a supremum it is a worst case concept. As we've seen, in IBC it is natural to consider average behavior. Werschulz places a measure on the inputs, e.g., on the initial conditions. He says a problem is well-posed on the average if the expectation with respect to this measure is finite. Then it can be shown that every ill-posed problem is well-posed on the average for every Gaussian measure, and hence solvable on the average; Werschulz and I \cite{4} survey the work leading to this result. We see that the unsolvability of ill-posed problems is a worst case phenomenon. It melts away on the average for reasonable measures.

I've tried to give you a taste of some of the achievements of IBC using the real-number model. The reader is invited to read some of the monographs cited in the References for other results and numerous open problems.

The research reported here was supported in part by the National Science Foundation and the Alfred P. Sloan Foundation. I appreciated the comments of Jerry B. Altzman, Anargyros Papageorgiou, and Lee Segel on the manuscript.

\end{document}